# Participatory Design to build better contact- and proximity-tracing apps

## Trust: the critical pillar of society

Abhishek Gupta*[+1,2] and Tania De Gasperis*[+3,4]

[1] Founder, Montreal AI Ethics Institute

[2] Machine Learning Engineer, Microsoft

[3] AI Ethics Researcher, Montreal AI Ethics Institute

[4] Strategic Innovation Lab (sLab), OCAD University

[+]Equal contributions

*Corresponding authors: Montreal, Canada (abhishek@montrealethics.ai) and (tania@montrealethics.ai)

**Abstract:**

With the push for contact- and proximity-tracing solutions as a means to manage the spread of the pandemic, there is a distrust between the citizens and authorities that are deploying these solutions. The efficacy of the solutions relies on meeting a minimum uptake threshold which is hitting a barrier because of a lack of trust and transparency in how these solutions are being developed. We propose participatory design as a mechanism to evoke trust and explore how it might be applied to co-create technological solutions that not only meet the needs of the users better but also expand their reach to underserved and high-risk communities. We also highlight the role of the *bazaar* model of development and complement that with quantitative and qualitative metrics for evaluating the solutions and convincing policymakers and other stakeholders in the value of this approach with empirical evidence.

**Data availability statement:**

Data sharing is not applicable to this article as no new data were created or analyzed in this study.

**Introduction**

Contact Tracing (CT) is being touted as *the* technological intervention that will help us "flatten the curve'' (McCurry, 2020) and manage COVID-19 cases so that we have more precise testing that judiciously utilizes our scarce medical and testing infrastructure. Each country is offering its own version of a CT application (Meixner, 2020), with some countries offering multiple solutions (Alawadhi, 2020). But, the success of these technologies lies in widespread adoption and so far the results have been mixed. Most countries continue to have low adoption rates (Aravindan & Phartiyal, 2020; "Contact-tracing apps enjoy", 2020), because of voluntary enrolment and without a significant percentage of the population downloading and utilizing the application, its efficacy comes into question. This stands in stark contrast to the public interest in wanting to use these solutions to improve public health outcomes (Milsom et al., 2020). So what is causing the divergence between expressing an interest in utilizing this technology and not following through in practice?

**Trust and the Social Fabric**

Trust is a central element underpinning the functioning of our society (Cook, 2001) and when we require mass-level coordination across cultures, geographies and other constructs, as is the case with the adoption of a CT solution, it becomes crucial that voluntary utilization emerges organically, especially in places where it is hard to enforce adoption in a top-down manner. With a technology that has the potential to track movements and monitor one's comings and goings on a 24/7 basis, it is not hard to imagine that this evokes an Orweillan nightmare that makes people uneasy. Privacy activists have rightly raised concerns on how the data will be collected, where it will be stored, who will have access to it, for how long will it be stored, what other data will it be associated with, among many other concerns. In places like South Korea ( "At a love motel", 2020), despite the anonymity of alerts provided to users of CT applications, people are able to reverse engineer identities, and it has led to ostracization and harm not only to the people directly involved, but also their families and the owners of businesses and organizations where they have been. Location data even when anonymized has the potential to reveal the identity of individuals (Gambs et al., 2014) if sufficient points are made available. In a recent incident in Morocco (Alami, 2020), the sexual orientation of several individuals was unveiled because of the combined actions of people responding to a call from a known activist who tried to spark a movement without fully realizing the implications that it would have, which highlights the challenges with ascertaining the unintended consequences when location data is collected *en masse*.

On the one hand, we have examples that show that people are apathetic to privacy policies (Bechmann, 2014) and are willing to give up personal information in the interest of using a product or service that they really desire. Reading through policies and understanding impacts for the services that we use on a daily basis requires weeks (McDonald & Cranor, 2008) of our time if they were to be gone through in detail and decisions to use them were made after proper analysis. When buried under legalese, the framing of the potential impacts of use of technology takes on an adversarial frame between the creators and the users of the application. The techlash (Hemphill, 2019) has been one significant consequence of this. When we have solutions that are being developed by large technology companies that have thus far evaded responsible innovation, it is natural to have hesitation on the part of the users in installing an application that will be pervasive in scope in terms of capturing data. Additionally, there is a rising concern that some solutions are being rushed to market with minimal regulatory oversight as a rapid response in trying to contain the spread of the pandemic (Mogensen, 2020).

There are a variety of other concerns including the explanation of the differences between location-based tracing (Wang & Loui, 2009) and proximity tracing (Cho et al., 2020) which can allay some of the concerns that citizens have in terms of what implications this is going to have on them. Creating accessible explanations (Tintarev & Masthoff, 2007) that do not rely heavily on technical jargon has the potential to increase trust and surface points of friction that can be included in the design and implementation process to build solutions that work for everyone, for example, by including language translations and being sensitive to cultural differences.

With many people using online technology for the first time (Finn, 2020), a change that has been accelerated by the pandemic ("The Covid-19 pandemic", 2020), the articulation of costs and benefits of using such technology and easing the process of installing, setting up, and managing new applications has gained even more importance. This is where we see inclusive and participatory design (PD) playing a key role in enabling us to meet the minimum threshold required for a solution like CT to be effective (Centers for Disease Control and Prevention, 2020). There are numerous examples (Inclusive Design Research Centre, 2020; Jacobs,1999) of how products and services when designed with the needs of everyone in mind help to spur innovation that really pushes the envelope in terms of what we can achieve with technology. Jutta Treviranus (2020) of the Inclusive Design Research Centre, states "people who have difficulty with or can't use our current systems are needed to move us forward, most innovations we take for granted today were catalyzed by the desire to circumvent a barrier experienced due to a disability."

## **Participatory Design**

We believe that an approach to combat the insufficient uptake of CT applications can be found in a value-centered design approach, such as PD. PD originated in Scandinavia in the 1970s (Bodker & Pekkola, 2010) and offers a unique opportunity due to its democratic nature and emphasis on collective shaping of a better future (Van der Velden et al., 2014). PD has been applied across research and industries in a variety of ways using multiple titles such as co-design, co-creation, cooperative design and design thinking, but the shared principle is that "research is not done *on* people as passive subjects providing *data* but *with* them to provide relevant information for improving their lives. The entire research process is viewed as a partnership between stakeholders…" (International Collaboration for Participatory Health Research, Position Paper 1, 2013).

What makes PD, co-design and its other titles unique, is that it includes all stakeholders of an issue, not just the users, and utilizes the input across the entire research and implementation process (Szebeko & Tan, 2010). Solutions thus far for CT applications have been predominantly built by deeply technical communities (Troncoso, 2020) in isolation which fail to solicit and utilize vast participant input in a co-design process. PD, because of its emphasis for located accountabilities in technology design (Suchman, 2002), encourages the process and legitimacy of having those affected by new technology be involved in its design (Kensing & Blomberg, 1998).

According to Kensing and Greenbaum, PD's guiding principles include: situation-based actions, mutual learning, tools and techniques, and alternative visions about technology, equalising power and democratic practices (Kensing & Greenbaum, 2012). The expertise and contextual relevance (International Collaboration for Participatory Health Research, Position Paper 2, 2013) that each individual brings to the PD process is valued (Onwuegbuzie et al., 2009), and there is a shared commitment to achieve outcomes that directly benefit those involved. From the Participatory Health Research perspective, this diverse expertise increases the relevance and uptake of outcomes, improves the health status and behaviours, and increases empowerment (Anderson et al., 2015; Cyril et al., 2015; O'Mara-Eves et al., 2015) serving as an example for the power of this approach and how it can improve uptake and use of contact- and proximity-tracing applications.

PD offers a range of methods to surface key results, such as prototyping, storyboards, future workshops (Van der Velden et al., 2014) which can be enacted in communities across a nation prior and during deployment of a CT application to encourage adoption and long-term sustained use (for the length of necessary time the CT application will be deployed).

PD can address some of the shortcomings of current CT applications uptake in many ways. PD can solidify trust through the equalising of power. As PD recognizes the lived experiences of participants, it is this recognition and mutual goal of creating a useful outcome that can aid in the empowerment of communities (Wallerstein, 2006), "it is this constructive and creative attention to people's aspirations and values, the political contexts, and the particularities of each setting that creates the depth and qualities of participatory design" (Smith et al., 2017). This can address the overarching concern of trust and create long-term and sustained behavioural change which will be critical in managing the negative effects of the pandemic. The actions of the PD process, including information sharing and shared problem identification activities also enhance trust (Selin et al., 2007).  For example, with those who are coming online for the first time, often the elderly, it is critical that they use these solutions as intended for their own safety; as is the case with the elderly, they are particularly susceptible to COVID-19 (Wallerstein, 2006) and being alerted about having been in contact with someone who has tested positive can mean the difference between life and death for them.

PD can also assist with uptake of CT applications as it helps us understand distrust. PD explores the hopes, fears and concerns of communities through social learning, defined as "learning that occurs when people deliberately engage [with] each other, sharing diverse perspectives and experiences to develop a common frame of understanding and basis for joint action" (Schusler et al., 2003) and mutual learning, as participants share their practical knowledge, "they also learn more about their work themselves" (Karasti, 2001). When you have a solution that has been developed in this manner, you create vocal community champions who, through the trusted referral process (Dobele & Lindgreen, 2011), rally the community behind them in adopting the solution.

From an economic perspective, PD improves the overall outcome as it includes diverse participants. This allows for more consideration of outliers and can create a more robust product that is accessible and usable by more people. It allows for proactive and preventative problem solving and can greatly eliminate the need for patching solutions later on that may open attack surfaces and potentially cause privacy breaches. Since PD's methods enable participants to "anticipate future use and alternative futures" (Van der Velden et al., 2014), there is the ability to catch problems earlier, which additionally helps to lower overall costs.

## Building software in the *bazaar*

Borrowing the metaphor of the *bazaar* and the cathedral model of software development (Raymond, 1999), the recent departure of experts ("PEPP-PT vs DP-3T", 2020) from the Pan-European Privacy Preserving Proximity Tracing (PEPP-PT) solution highlights how a non-transparent, centralized approach fails when it clashes with a solution that is built using the bazaar model such as the DP3T (Troncoso, 2020) framework.

The *bazaar* approach embodies the Open Source Software (OSS) movement which relies on the ideas of openness in development while inviting anyone to collaboratively build on top of existing work. This approach has demonstrably created more secure and robust solutions (Craig-Wood, 2013) and as a consequence evokes higher levels of trust (Petrinja et al., 2008) from the users. Other solutions that are being proposed at the moment, including the proposal from Apple and Google ("Apple and Google partner", 2020) suffer from a lack of transparency, not in the design specifications which are available publicly ("Privacy-Preserving Contact,", 2020), but in the documentation of how those design decisions were made, who made them, which alternatives were evaluated and why some were chosen over others. A git ("About - Git," 2020) like history-tracking of the evolution of the design choices and specific tradeoffs bring true transparency (Ram, 2013) which is key in getting users to trust that the solutions being proposed are being built with no malicious, subversive intentions. It also provides a clear documentation of the work-in-progress history that showcases how the solution is evolving and that open participation is not only appreciated but encouraged.

From a PD perspective, we can obtain higher levels of success when people know that their inputs were valued and included, which makes them vocal champions for the adoption of that solution; this serves two purposes: one being that potential problems are caught early which makes it easier to fix them (both in terms of time and resources) and two being that uptake of solutions is higher because often these active participants are well-recognized community leaders who are responsible for swaying technology adoption decisions. We also want to avoid the situation of ethics-washing (Wagner, 2018) where the PD process might be done as a token exercise to check off a box but inputs are not really incorporated. This problem is nixed by using a radical transparency (Birchall, 2014) and meaningful transparency ("Mission, Team and Story," 2020) approach where such ethics-washing could be easily called out. Greater behavior change can be triggered when participants in the PD process know that their inputs are well-represented, valued and considered in the design and creation of a solution. The behavior change is also more long-term and sustained (Ssozi-Mugarura, 2017) which is crucial in the case of contact- and proximity-tracing applications for them to be effective in curbing the spread of the pandemic.

**Proposed indicators to measure success:**

Coupling this qualitative approach with quantitative indicators will be crucial in convincing both the policymakers and the institutions that are developing CT solutions. To that end, we propose utilizing the following indicators as concrete metrics to measure the effectiveness of the PD perspective in increasing uptake and consistent utilization of CT solutions:

1. Split-demographic information to track how high-risk and hitherto underserved populations are using the application
2. Time-period oriented growth rates measured across geographies, demographics and other variables of interest
3. Tracking sustained use through Daily Active Users, Weekly Active Users, and other measures coupled with cohort analysis that indicate consistent use of the application

We recognize that some of the proposed metrics above are proxies for measuring the real impact, yet they provide intelligence that will be useful in making a case for the use of PD methods. The PD process itself can be utilized to create and tailor metrics that better meet the contextual and cultural needs of different subpopulations.

In addition, the following qualitative indicators from the PD perspective will supplement the insights gleaned from the quantitative measures:

1. Improvement of the quality of life (Spinuzzi, 2005) - the solution, such as a CT application, must have a direct benefit on the participants and their community.
2. Continuous checks (Spinuzzi, 2005) - the solution will require continuous checks with community members to assess any obstacles or newly formed requirements that should be addressed and improved on. This iterative process provides a robust and adaptive approach to ensure sustained use of the solution.
3. Effectiveness of the solution (Drain et al., 2018) - the solution must meet the needs of the participants and their community and meet the original requirements of the project.
4. Likelihood of adoption (Drain et al., 2018) - the solution should be adopted by the community.

**Future Research Areas:**

The PD perspective lays the groundwork for building successful co-owned and co-developed solutions. With the rapid development and deployment of CT solutions, we encourage other researchers to join us in our efforts to build a better understanding in the following areas which will be crucial in enhancing the efficacy of the PD approach:

1. Effective PD during isolation and social distancing - this includes gaining a better understanding of how to effectively leverage online platforms to run co-design sessions which are typically held in-person that allow for serendipitous interactions to organically emerge in a physical environment. This is particularly constrained in simultaneous video conferencing which largely only permits one-to-many communication because of lags in transmission and limited ability to have side-channel conversations.

2. Accessible technical communication design - to achieve transparency and trust in technical solutions, the wider audience must not only be able to easily find and access the specifications for various technical solutions but they must also be intelligible where intelligibility can be defined as per the ontology proposed in the machine learning literature (Zhou & Danks, 2020): intelligibility for Engineers, Users and Affectees.

3. Version tracking of design choices such that there is a historical record that is available to the relevant stakeholders so they can make determinations about the evolution of intentions and avoid things like function creep (Tzanou, 2010).

4. Sandboxing and phased roll-outs of CT solutions so that they are tested early and frequently to catch issues and ensure that the needs of different cultures and populations are met appropriately.

**Conclusion**

In this paper, we have explored how current CT solutions for COVID-19 are facing barriers in uptake and we have proposed how trust and transparency, rooted in social psychology can act as the expedients for greater and sustained adoption of these solutions. PD is put forth as a perspective that helps to achieve these goals of trust and transparency while simultaneously achieving the goal of building more inclusive solutions that bring traditionally marginalized communities into the fold. In addition, adopting the *bazaar* model of development, borrowing from the software development ideology coupled with the mentioned quantitative and qualitative measures will help to convince stakeholders of the value in this approach by grounding it in empirical evidence. We conclude with potential research areas that can continue to build on the approach mentioned here so that we can emerge from the crisis of the pandemic without a crisis of distrust if CT and other technologies are deployed to manage the current harms.